# Quasi-linear buildup of Coulomb integrals via the coupling strength parameter in the non-relativistic electronic Schrödinger equation

Sandor Kristyan

*Research Centre for Natural Sciences, Hungarian Academy of Sciences,*
*Institute of Materials and Environmental Chemistry*
*Magyar tudósok körútja 2, Budapest H-1117, Hungary,*

Corresponding author: <u>kristyan.sandor@ttk.mta.hu</u>

**Abstract**. The non-relativistic electronic Hamiltonian, $H_\nabla + H_{ne} + aH_{ee}$, is linear in coupling strength parameter (a), but its eigenvalues (electronic energies) have only quasi-linear dependence on it. Detailed analysis is given on the participation of electron-electron repulsion energy ($V_{ee}$) in total electronic energy ($E_{total\ electr,k}$) in addition to the well-known virial theorem and standard algorithm for $v_{ee}(a=1)=V_{ee}$ calculated during the standard- and post HF-SCF routines. Using a particular modification in the SCF part of the Gaussian package, we have analyzed the ground state solutions via the parameter "a". Technically, with a single line in the SCF algorithm, operator was changed as $1/r_{ij} \to a/r_{ij}$ with input "a". The most important findings are, 1, $v_{ee}(a)$ is quasi-linear function of "a", 2, the extension of 1st Hohenberg-Kohn theorem ($\Psi_0(a=1) \Leftrightarrow H_{ne} \Leftrightarrow Y_0(a=0)$) and its consequences in relation to "a". The latter allows an algebraic transfer from the simpler solution of case a=0 (where the single Slater determinant $Y_0$ is the accurate form) to the physical case a=1. Moreover, we have generalized the emblematic Hund's rule, virial-, Hohenberg-Kohn- and Koopmans theorems in relation to the coupling strength parameter.

## INTRODUCTION

The non-relativistic, spinless, fixed nuclear coordinate electronic Schrödinger equation (SE) for molecular systems containing M atoms and N electrons with nuclear configuration $\{\mathbf{R}_A, Z_A\}_{A=1}^M$ in free space is

$$H(a)\ y_k(a) \equiv (H_\nabla + H_{ne} + aH_{ee})\ y_k(a) = enrg_{electr,k}(a)\ y_k(a) \ . \tag{1}$$

Energy operators are the kinetic, $H_\nabla \equiv -\Sigma_{i=1}^N \nabla_i^2/2$, nuclear–electron attraction, $H_{ne} \equiv -\Sigma_{i=1}^N \Sigma_{A=1}^M Z_A R_{Ai}^{-1}$, and electron–electron repulsion, $H_{ee} \equiv \Sigma_{i=1}^N \Sigma_{j=i+1}^N r_{ij}^{-1}$. The $(y_k(a), enrg_{electr,k}(a))$ is the k-th eigenvalue pair of electronic Hamiltonian H(a), for which we use notations $(Y_k, e_{electr,k})$ if a=0 and $(\Psi_k, E_{electr,k})$ if a=1. The $y_k(a)$ is anti-symmetric, well behaving, normalized as $\langle y_k|y_k\rangle=1$ and the one-electron density is $\rho_k(\mathbf{r}_1,a) = N\int y_k^* y_k ds_1 d\mathbf{x}_2 \ldots d\mathbf{x}_N$. $S_0$ (generally $s_0(a)$) is a single determinant approximation for $\Psi_0$ (a=1, generally for $y_0(a)$) via HF-SCF/basis/a energy minimizing algorithm, $enrg_{electr,k}(a) \leq enrg_{electr,k+1}(a)$ for k=0,1,2,... .The ground state (k=0) two-electron density is $b_0(\mathbf{r}_1,\mathbf{r}_2,a) = (N(N-1)/2)\int y_0^* y_0 ds_1 ds_2 d\mathbf{x}_3 \ldots d\mathbf{x}_N$ providing $\int b_0(\mathbf{r}_1,\mathbf{r}_2,a) d\mathbf{r}_1 d\mathbf{r}_2 = ((N-1)/2)\int \rho_0(\mathbf{r}_1,a) d\mathbf{r}_1 = N(N-1)/2$. ($\int b_0(\mathbf{r}_1,\mathbf{r}_2,a) d\mathbf{r}_1 d\mathbf{r}_2 =$ number of electron pairs, and $\int \rho_0(\mathbf{r}_1,a) d\mathbf{r}_1 = N=$ number of electrons for any value of a.) <u>The $y_k(a=0)=Y_k$ has a single determinant form, while $y_k(a\neq0)$ does not</u>. $E_{electr,0}$(method) approximates $E_{electr,0}$ by a certain method: Hartree-Fock Self Consistent Field (HF-SCF) and Kohn-Sham (KS) of density functional theory (DFT) for the vicinity of stationary (equilibrium or transition state) points, configurations interactions (CI) for any nuclear geometry, etc.. The a=1 is the physical (real) case, where $\Psi_k$ and $E_{electr,k}$ are the k$^{th}$ excited state (k=0,1,2,...) anti-symmetric wave function (with respect to all spin-orbit electronic coordinates $\mathbf{x}_i \equiv (\mathbf{r}_i, s_i)$) and electronic energy, resp.; $E_{total\ electr,k} = E_{electr,k} + V_{nn}$, where $V_{nn} \equiv \Sigma_{A=1}^M \Sigma_{B=A+1}^M Z_A Z_B/R_{AB}$ is the nuclear repulsion term. Case a=0 mathematically provides a good starting point to solve the very important problem when a=1. The physical $V_{ee} \equiv v_{ee}(a=1) = \langle \Psi_0 | H_{ee} | \Psi_0 \rangle$ where $v_{ee}(a) \equiv \Sigma_{i=1}^N \Sigma_{j=i+1}^N \int y_0^* y_0 (a/r_{ij}) d\mathbf{x}_1 d\mathbf{x}_2 \ldots d\mathbf{x}_N =$ $a(N(N-1)/2) <y_0|r_{12}^{-1}|y_0> = a<y_0|H_{ee}|y_0>$ with normalized $y_0$. The $a<0$ would mean "attractive electrons". The dimensionless coupling strength parameter "a" scales the electron-electron interaction energy, $v_{ee}(a)$. For example, "a" is capable [1-2] to correct the HF-SCF energy remarkably well with scaling "a" a bit below unity, as well as it defines the adiabatic connection (AC) in KS formalism. The kinetic energy is $T \equiv <\Psi_0|H_\nabla|\Psi_0>$ and the electron-nuclear attraction is $V_{ne} \equiv <\Psi_0|H_{ne}|\Psi_0>$.

HF-SCF/basis/a=1 procedure minimizes the functional $<S_0|H|S_0>$ ($> <\Psi_0|H|\Psi_0> \equiv E_{electr,0} = T+V_{ne}+V_{ee}$) for a normalized single Slater determinant approximate wave function ($S_0$) with constrained (ortho-normalized) molecular orbitals (MO). The $\approx 1\%$ error of $<S_0|H|S_0>$ (with at least minimal basis set at near stationary points) known as correlation energy ($E_{corr}$) is negative by the variation principle (VP), necessary for chemical accuracy (CA, 1 kcal/mol).

The HF-SCF/basis/a=0 algorithm solves $(H_\nabla + H_{ne})Y_k = e_{electr,k} Y_k$ for ground- (k=0) and excited states (k>0, see trick below), with Slater determinant (correct form) for $Y_k$ with numerical (basis set) error and $E_{corr}(a=0,k\geq 0)=0$; we notate energies as $t \equiv <Y_0|H_\nabla|Y_0>$ and $v_{ne} \equiv <Y_0|H_{ne}|Y_0>$. The HF-SCF/basis/a=1 or a≠0 provides $S_0$ or $s_0(a)$ determinant (incorrect form) for $\Psi_0$ or $y_0(a)$ with $E_{corr}(a \neq 0, k=0)<0$ and numerical (basis set) error, as well as for k>0 the estimation is very weak. Importantly, the $S_0$ is very close to $Y_0$; $\lim_{a \to 0 \text{ or } 1} s_0(a) = Y_0$ or $S_0$. Moreover, calculating $Y_0$ in this way is not restricted to the vicinity of the stationary point, while if a≠0 it does. The HF-SCF/basis/a=0 calculation is faster, more stable and less memory taxing in comparison to HF-SCF/basis/a=1. Recall the well-known device in the theory of ordinary diff. equations when one starts from the elementary homogeneous (e.g. y''+y=0) vs. non-homogeneous case (e.g. y''+y=f(x)): Here we use similar device for partial diff. eigenvalue Eq.1 via solving with a=0 (simpler task) to generate somehow the solution for a=1.

## CALCULATING GROUND STATE WITH a=0

**Totally non-interacting reference system (TNRS) is defined with a=0.** For any "a", Eq.1 is a linear partial diff. eigenvalue equation, the VP holds, the 1st ("$\rho_0(\mathbf{r}_1,a)$ defines $y_0$ and the nuclear frame") and 2nd ("VP for $\rho_0(\mathbf{r}_1,a)$ in the DFT functional") Hohenberg – Kohn (HK) theorems hold. Furthermore, the DFT functional $v_{ne}[\rho_{0,trial}] = -\Sigma_{A=1}^M Z_A \int R_{A1}^{-1} \rho_{0,trial}(\mathbf{r}_1,a) d\mathbf{r}_1$ is 100 % accurate form, while the kinetic DFT functional has the known difficulty for any "a". The energetically lowest lying eigenvalue pair ($e_{electr,0}, Y_0$) corresponds to ($E_{electr,0}, \Psi_0$), $E_{electr,0} >> e_{electr,0}$ for any molecular system (in stationer or non-stationer geometry, the large difference stems from the lack of $V_{ee}(a=1)$ when a=0). The ground state versus the energetically lowest lying state with an enforced spin multiplicity feature is also the same for any "a". However, if spin-spin interaction is not considered via Coulomb repulsion, Hund's rule does not apply for a=0 itself, but extension and approximation in e.g. Eq.20 below sets it back on the right track. On the other hand, there are major mathematical differences between a=1 and a=0: Operator $H_{ee}$ is very special in the sense that algebraically it is the "simplest" term, but in contrast it introduces the most difficult effect in HF-SCF, known as the non-classical Coulomb effect. $H_{ee}$ operator is responsible that a single Slater determinant $S_0$ for $\Psi_0$ (a=1) is not enough for total accuracy, although in the vicinity of stationary points it provides a good approximation, and it can provide many characteristic properties of the ground state eigenvalue, however, a=0 eliminates this problem.

If a=0, Eq.1 becomes $(H_\nabla + H_{ne})Y_k = \Sigma_{i=1}^N h_i = e_{electr,k} Y_k$, where $h_i \equiv -\nabla_i^2/2 - \Sigma_{A=1}^M Z_A R_{Ai}^{-1}$ is the one-electron operator. It decomposes to one-electron equations with linear dependence on "a" as

$$(h_i + aV_{ee,eff}(\mathbf{r}_i)) \phi_i(\mathbf{r}_i) = \varepsilon_i \phi_i(\mathbf{r}_i) \qquad (2)$$

where $\phi_i(\mathbf{r}_i)$ is the $i^{th}$ MO, and technically $\phi_i$ counts the MOs with the index i, so the notation is reducible from ($h_i, \phi_i(\mathbf{r}_i), \varepsilon_i$) to ($h_1, \phi_i(\mathbf{r}_1), \varepsilon_i$). $V_{ee,eff}$ is the effective potential from electron-electron repulsion; (other way [3] is shifting $H_{ne}$ algebraically into $V_{ee,eff}$, called $V_{eff}$). $V_{ee,eff}$ is expressed with the known J and K integrals in HF-SCF theory, or $V_{ee,eff}(\mathbf{r}_i) = \int \rho_0(\mathbf{r}_2,KS)r_{12}^{-1} d\mathbf{r}_2 + V_{xc}(\mathbf{r}_i)$ in KS formalism (the 1st term is the classical Coulomb term, the 2nd is the non-classical Coulomb term for "exchange-correlation"). $V_{ee,eff}(\mathbf{r}_i)$ is the term where the N equation in Eq.2 is coupled if a≠0. Eq.2 is coupled, though virtually not coupled, so the 100% adequate anti-symmetric solution for the equation system in Eq.2 (but not for a≠0 in Eq.1) is a Slater det, and this system is known as "non-interacting reference system" [3]. If even a=0 is set, a single Slater determinant as anti-symmetric solution is not only 100 % adequate for Eq.2, but also for a=0 in Eq.1, since all operators are one-electron operators, and we call a=0 in Eqs.1-2 as TNRS. More simply, Eq.2 should not be considered as an equation system containing N equations enumerated by i ($h_i \phi_i = \varepsilon_i \phi_i$), but in fact it is a single eigenvalue equation ($h_1 \phi_i = \varepsilon_i \phi_i$) providing ortho-normality $<\phi_i|\phi_j> = \delta_{ij}$, (in HF-SCF/bais/a≠0 the orto-normality is enforced). Eigenvalues of Eq.2 are ($\varepsilon_i$, $\phi_i(\mathbf{r}_1)$) for i=1,2,…∞, the i=1 is the lowest lying state and it is the lowest lying MO for a=0 in Eq.1 in its k=0 ground state. The single Slater determinant for a=0 in Eq.1 is accomplished for N electrons from the eigenvaues of Eq.2 ($Y_0=s_0(a=0)$), just as in the basic HF-SCF theory ($S_0=s_0(a=1)$). Eq.2 with the value of N and Eq.1 with a=0 are equivalent, more, it holds for symmetric and anti-symmetric $Y_k$ as well. If a=0 and e.g. $\phi_i$= f($\mathbf{r}_1$), g($\mathbf{r}_1$), h($\mathbf{r}_1$) are states of Eq.2 (MO's of Eq.1) with energies $\varepsilon_1 \leq \varepsilon_2 \leq \varepsilon_3$, resp. and N=3, some anti-symmetric eigenfunctions (wave functions) and eigenvalues (electronic energies) are $Y_0=|\alpha_1 f, \beta_2 f, \alpha_3 g>$ with $e_{electr,0}=2\varepsilon_1+\varepsilon_2$, $Y_{k'}=|\alpha_1 f, \alpha_2 g, \beta_3 g>$ with $e_{electr,k'}=\varepsilon_1+2\varepsilon_2$ and $Y_{k''}=|\alpha_1 f, \alpha_2 g, \alpha_3 h>$ with $e_{electr,k''}=\varepsilon_1+\varepsilon_2+\varepsilon_3$, etc.. The spin multiplicities are 2(1/2 –1/2 +1/2)+1=2, 2, 4, resp.. The electronic energy of the system in Eq.1 with a=0 is the sum of energy levels, generally speaking weighted as populated:

$$\{e_{electr,k}= \Sigma_i n_i \varepsilon_i \text{ if } a=0\} \quad \text{and} \quad \{enrg_{electr,k}(\text{HF-SCF/basis/a}) \neq \Sigma_i n_i \varepsilon_i \text{ if } a \neq 0\}, \quad (3)$$

where $n_i$ is the population of the i[th] energy level: 0, 1 or 2, the lattermost is with opposite spins. Also, $\Sigma_{i=1} n_i = N$. Eq.3 between MO energies ($\varepsilon_i$) and ground state electronic energy ($e_{electr,0}$) holds via HF-SCF/basis/a=0, but not via (HF-SCF or KS)/basis/a≠0, in the latter some cross terms must be subtracted [4] (that goes to zero if a→0). The definition of restricted (RHF) and unrestricted (UHF) form of Slater determinants also lose their necessity if a=0. For a value of N and multiplicity 2S+1= 2$\Sigma s_i$+1 in the regular way, the HF-SCF/basis/a=0 algorithm calculates the lowest lying N/2 or (N+1)/2 energy values ($\varepsilon_i$) and MOs ($\phi_i$), the latter with linear combination of atomic orbitals (LCAO) expansion of the basis set for Eq.1 with a=0 or Eq.2.

**Links between a=1 and a=0 (TNRS) in relation to ground state** comes from $<\Psi_0|H_{ee}|Y_0> = <\Psi_0|H - (H_\nabla+H_{ne})|Y_0> = <\Psi_0|H|Y_0> - <\Psi_0|(H_\nabla+H_{ne})|Y_0> = <Y_0|H|\Psi_0> - e_{electr,0}<\Psi_0|Y_0> = E_{electr,0}<Y_0|\Psi_0> - e_{electr,0}<\Psi_0|Y_0>$ by the hermetic property. Finally, with $\xi \equiv <\Psi_0|r_{12}^{-1}|Y_0>/<\Psi_0|Y_0>$ one obtains

$$E_{electr,0}= e_{electr,0} + <\Psi_0|H_{ee}|Y_0>/<\Psi_0|Y_0>= e_{electr,0} + (N(N-1)/2)\xi . \quad (4)$$

The $\xi \approx$ 0.3-0.5 h and ratio ($E_{electr,0} - e_{electr,0})/e_{electr,0} \approx$ -0.4 as functions of molecular frame ($H_{ne}$) are <u>quasi-constants</u>. Compare to the <u>exactly constant</u> value 2 in virial theorem: $(V_{nn}+V_{ne}+V_{ee})/T = –2 = (V_{nn}+v_{ne})/t \equiv (V_{nn}+ <Y_0|H_{ne}|Y_0>)/<Y_0|H_\nabla|Y_0>$ holding for atoms, atomic ions and equilibrium/transition state geometry molecules, i.e. the value 2 is invariant on $H_{ne}$. There is significant difference between $V_{ee} \equiv <\Psi_0|H_{ee}|\Psi_0>= (N(N-1)/2)<\Psi_0|r_{12}^{-1}|\Psi_0>$ and the corresponding $<\Psi_0|H_{ee}|Y_0>/<\Psi_0|Y_0>= (N(N-1)/2)<\Psi_0|r_{12}^{-1}|Y_0>/<\Psi_0|Y_0>$. The divisor $<\Psi_0|\Psi_0>$ comes up in $V_{ee}$ if it is not normalized to unity, making the algebraic analogy even closer between $V_{ee}$ and $(N(N-1)/2)\xi$. $V_{nn}$ is cancelled in $E_{total\ electr,0} - e_{total\ electr,0}= E_{electr,0} - e_{electr,0}$. A more general expression between k and k' excited states, coming from the same derivation as

$$E_{electr,k}= e_{electr,k'} + (N(N-1)/2)<\Psi_k|r_{12}^{-1}|Y_{k'}>/<\Psi_k|Y_{k'}> \quad (5)$$

and Eqs.4-5 forecast the generalization of 1[st] HK theorem, detailed later.

Obviously $E_{electr,k} >> e_{electr,k}$ by the $1/r_{ij} \geq 0$, recall CA. From the VP, let the normalized $Y_0$ be a trial for a=1, and one gets $E_{electr,0} \leq <Y_0|H|Y_0> = <Y_0|H_{ee}|Y_0> + <Y_0|H_\nabla+H_{ne}|Y_0> = <Y_0|H_{ee}|Y_0> + e_{electr,0}$, that is

$$E_{electr,0} \leq e_{electr,0} + <Y_0|H_{ee}|Y_0> . \quad (6)$$

The reverse situation, when $\Psi_0$, the solution for a=1 is a trial function for a=0, one gets the simpler

$$e_{electr,0} \leq <\Psi_0|H_\nabla+H_{ne}|\Psi_0> . \quad (7)$$

In trivial case, $\Psi_0(N=1)= Y_0(N=1)$. Via a=1, it separates as $<\Psi_0|H|\Psi_0> = <\Psi_0|H_\nabla+H_{ne}|\Psi_0> + <\Psi_0|H_{ee}|\Psi_0> = E_{electr,0}$, and the right hand side is majored by Eq.6 as $<\Psi_0|H_\nabla+H_{ne}|\Psi_0> + <\Psi_0|H_{ee}|\Psi_0> \leq e_{electr,0} + <Y_0|H_{ee}|Y_0>$, and with Eq.7 one obtains

$$<\Psi_0|H_{ee}|\Psi_0> \leq <Y_0|H_{ee}|Y_0> . \quad (8)$$

The counterpart of Eq.6 comes from Eq.7 as $e_{electr,0} + <\Psi_0|H_{ee}|\Psi_0> \leq <\Psi_0|H_\nabla+H_{ne}|\Psi_0> + <\Psi_0|H_{ee}|\Psi_0> = <\Psi_0|H|\Psi_0> = E_{electr,0}$ which is

$$E_{electr,0} \geq e_{electr,0} + <\Psi_0|H_{ee}|\Psi_0> . \quad (9)$$

$$e_{electr,0} << (e_{electr,0}+<\Psi_0|H_{ee}|\Psi_0>) \leq E_{electr,0}= (e_{electr,0}+<\Psi_0|H_{ee}|Y_0>/<\Psi_0|Y_0>) \leq (e_{electr,0}+<Y_0|H_{ee}|Y_0>) \quad (10)$$

is the full relation in summary, which extends Eq.8 as

$$<\Psi_0|H_{ee}|\Psi_0>) \leq <\Psi_0|H_{ee}|Y_0>/<\Psi_0|Y_0> \leq <Y_0|H_{ee}|Y_0> . \quad (11)$$

The relationships in Eq.4 to 10 can be developed further with the Hellmann–Feynman theorem [5] $\partial E_{electr,k}/\partial\lambda = <\Psi_k|\partial H(\lambda)/\partial\lambda|\Psi_k>$ with normalization $<\Psi_k|\Psi_k>=1$ and $\lambda:=a$. Using the anti-symmetric property and $\partial H(a)/\partial a = \partial(aH_{ee})/\partial a= H_{ee}$ for ground state (k=0)

$$\partial enrg_{electr,0}(a)/\partial a= <y_0|H_{ee}|y_0>= (N(N-1)/2) <y_0|r_{12}^{-1}|y_0> \quad (12)$$

$$E_{electr,0} - e_{electr,0}= (N(N-1)/2) \int_{[0,1]}<y_0|r_{12}^{-1}|y_0>da \quad (13)$$

$$\int_{[0,1]}<y_0|r_{12}^{-1}|y_0>da= <\Psi_0|r_{12}^{-1}|Y_0>/<\Psi_0|Y_0> \quad (14)$$

where the interval [0,1] can be extended to a general $[a_1,a_2]$. (The $y_0(a)$ is normalized for all "a", but $\langle\Psi_0|Y_0\rangle$ in the denominator in Eq.4 or 14 is not necessarily unity.) The right hand side of Eq.12 is

$$\partial \text{enrg}_{electr,0}(a)/\partial a = \int b_0(\mathbf{r}_1,\mathbf{r}_2,a)\, r_{12}^{-1} d\mathbf{r}_1 d\mathbf{r}_2 = v_{ee}(a)/a \,, \qquad (15)$$

where the linear multiplier "a" in $v_{ee}$ has disappeared, it has only effect from inside the normalized $y_0$. The $a=0$ cancels $H_{ee}$, but the non-vanishing value $v_{ee}(a)/a$ is present. Via Eq.15, Eq.13 is alternatively

$$E_{electr,0} - e_{electr,0} = \int_{[0,1]} \int b_0(\mathbf{r}_1,\mathbf{r}_2,a)\, r_{12}^{-1} d\mathbf{r}_1 d\mathbf{r}_2 da \,. \qquad (16)$$

The classical approximation in KS [5] for exchange-correlation and self-interaction yields $\partial \text{enrg}_{electr,0}(a)/\partial a \approx (1/2)\int \rho_0(\mathbf{r}_1,a)\rho_0(\mathbf{r}_2,a) r_{12}^{-1} d\mathbf{r}_1 d\mathbf{r}_2$ and $E_{electr,0} - e_{electr,0} \approx (1/2)\int_{[0,1]} \int \rho_0(\mathbf{r}_1,a)\rho_0(\mathbf{r}_2,a) r_{12}^{-1} d\mathbf{r}_1 d\mathbf{r}_2 da$. Eqs.15-16 do not have exchange-correlation effect by the use of $b_0$ [6-7]. The 2nd derivative from Eq.12 for real $y_0$ is

$$\partial^2 \text{enrg}_{electr,0}(a)/\partial a^2 = N(N-1)\langle y_0|r_{12}^{-1}|\partial y_0/\partial a\rangle = \int (\partial b_0(\mathbf{r}_1,\mathbf{r}_2,a)/\partial a) r_{12}^{-1} d\mathbf{r}_1 d\mathbf{r}_2 = (\partial/\partial a)\int b_0(\mathbf{r}_1,\mathbf{r}_2,a) r_{12}^{-1} d\mathbf{r}_1 d\mathbf{r}_2. \qquad (17)$$

Importantly, $\partial \text{enrg}_{electr,0}(a)/\partial a \approx$ const., more, $\text{enrg}_{electr,0}(a)$ is quasi-linear with simple curvature and less linear with increasing basis set (Fig.1). The $\partial \text{enrg}_{electr,0}(a)/\partial a$ is a quasi-constant $\Rightarrow$ "linear approximation" is possible (Eq.4$\rightarrow$Eq.20), as well as the <u>LCAO coefficients are close to each other between $Y_0$ and $\Psi_0$</u>. Generalization of Eq.4 comes from $\langle y_0(a_2)|(a_2-a_1)H_{ee}|y_0(a_1)\rangle = \langle y_0(a_2)|H(a_2)-H(a_1)|y_0(a_1)\rangle = \langle y_0(a_2)|H(a_2)|y_0(a_1)\rangle - \langle y_0(a_2)|H(a_1)|y_0(a_1)\rangle = (\text{enrg}_{electr,0}(a_2) - \text{enrg}_{electr,0}(a_1))\langle y_0(a_2)|y_0(a_1)\rangle$, from which

$$\text{enrg}_{electr,0}(a_2) = \text{enrg}_{electr,0}(a_1) + (a_2-a_1)(N(N-1)/2)\langle y_0(a_2)|r_{12}^{-1}|y_0(a_1)\rangle/\langle y_0(a_2)|y_0(a_1)\rangle \qquad (18)$$

with $\xi(a_1,a_2) \equiv \langle y_0(a_2)|(a_2-a_1)r_{12}^{-1}|y_0(a_1)\rangle/\langle y_0(a_2)|y_0(a_1)\rangle$ and similarly for excited states. If $a_1=0$ and $a_2=a$

$$\text{enrg}_{electr,0}(a) = e_{electr,0} + (N(N-1)/2)\langle y_0|ar_{12}^{-1}|Y_0\rangle/\langle y_0|Y_0\rangle \,. \qquad (19)$$

The $\lim_{a_1\rightarrow a_2=a}$ in Eq.18 yields Eqs.12 or 15. Although $v_{ee}(a=0)=0$ is trivial, but $(v_{ee}(a)/a)|_{a=0} \neq 0 \Rightarrow \partial \text{enrg}_{electr,0}(a)/\partial a|_{a=0} = (N(N-1)/2)\langle Y_0|r_{12}^{-1}|Y_0\rangle \neq 0$ in Eqs.12,15 (Fig.1). On Fig.1 the relative $(\text{enrg}_{electr,0}(a) - e_{electr,0})/e_{electr,0} = \langle y_0|aH_{ee}|Y_0\rangle/(\langle y_0|Y_0\rangle\langle Y_0|H_\nabla+H_{ne}|Y_0\rangle)$ value from Eq.19 is plotted, wherein $y_0(a)$ is approximated with $s_0(a)$ from a HF-SCF/basis/a algorithm, the slope at $a=0$ is $\partial((\text{enrg}_{electr,0}(a) - e_{electr,0})/e_{electr,0})/\partial a|_{a=0} = (\partial \text{enrg}_{electr,0}(a)/\partial a)|_{a=0}/e_{electr,0} = \langle Y_0|H_{ee}|Y_0\rangle/\langle Y_0|H_\nabla+H_{ne}|Y_0\rangle$ varying weakly with "a".

Eq.17 yields near zero value, because $1 = \langle y_0|y_0\rangle \Rightarrow 0 = \partial 1/\partial a = \partial\langle y_0|y_0\rangle/\partial a = 2\langle y_0|\partial y_0/\partial a\rangle \Rightarrow v_{ee}(a) \equiv a\langle y_0|H_{ee}|y_0\rangle$ is large, while $\langle y_0|r_{12}^{-1}|\partial y_0/\partial a\rangle$ in Eq.17 is small $\Rightarrow \partial^2 \text{enrg}_{electr,0}(a)/\partial a^2 \approx 0 \Rightarrow$ Eqs.18-19 are about straight lines with respect to "a", see Fig.1 $\Rightarrow$ Eq.19 with Eq.12 or 15 yields $\text{enrg}_{electr,0}(a) \approx e_{electr,0} + [(\partial \text{enrg}_{electr,0}(a)/\partial a)|_{a=0}]a = e_{electr,0} + a[(N(N-1)/2)\langle Y_0|r_{12}^{-1}|Y_0\rangle]$, and at $a=1$

$$E_{electr,0} \approx E_{electr,0}(\text{TNRS}) \equiv e_{electr,0} + (N(N-1)/2)\langle Y_0|r_{12}^{-1}|Y_0\rangle \,. \qquad (20)$$

More generally, the right side is $e_{electr,0} + (v_{ee}(a)/a)|_{a=0}$ for which Eq.10 gives the lower boundary. The last term in Eq.20 is $2J - K$, the known Coulomb- and exchange integrals well known in HF-SCF formalism for $S_0(a=1)$, here for $Y_0(a=0)$. KS approx. in Eq.20 yields $(N(N-1)/2)\langle Y_0|r_{12}^{-1}|Y_0\rangle \approx \frac{1}{2}\int \rho_0(\mathbf{r}_1,a=0)\rho_0(\mathbf{r}_2,a=0)r_{12}^{-1} d\mathbf{r}_1 d\mathbf{r}_2$, very generally, $(N(N-1)/2)\langle y_k(a)|r_{12}^{-1}|y_k(a)\rangle = \int b_k(\mathbf{r}_1,\mathbf{r}_2,a) r_{12}^{-1} d\mathbf{r}_1 d\mathbf{r}_2 \approx \frac{1}{2}\int \rho_k(\mathbf{r}_1,a)\rho_k(\mathbf{r}_2,a)r_{12}^{-1} d\mathbf{r}_1 d\mathbf{r}_2$. The latter suffers from exchange-correlation like Eq.20, (recall if one omits the term $r_{12}^{-1}$ (changing the repulsion op. to electron counting) one obtains $N(N-1)/2 \approx N^2/2$, indeed, not equal). From the VP, the $S_0$ from HF-SCF/basis/a=1 is energetically better than $Y_0$ from HF-SCF/basis/a=0 to use in Eq.1 with a=1, that is

$$E_{electr,0} < \langle S_0|H|S_0\rangle \leq e_{electr,0} + (N(N-1)/2)\langle Y_0|r_{12}^{-1}|Y_0\rangle \,. \qquad (21)$$

In Eq.21 (more general than Eq.6) equality may come up in "≤" when small e.g., STO-3G basis set is used (a fortunate coincidence). Finishing this section from a mathematical point of view, if an additional operator is in effect beside $H_{ee}$ referring to e.g., external forces, the algorithm or procedure is exactly the same as the one leading to Eq.20 and its discussion; the operator $H_{ee}$ must be changed or to be extended accordingly.

**Virial theorem at any value of "a", especially at a=1 and a=0 (TNRS)**, holds as

$$(V_{nn} + \langle y_0|H_{ne}|y_0\rangle + a\langle y_0|H_{ee}|y_0\rangle)/\langle y_0|H_\nabla|y_0\rangle = -2 \qquad (22)$$

for atoms ($V_{nn}=0$) and stationary molecules. For a=1, it is $(V_{nn} + V_{ne} + V_{ee})/T = -2$, while for a=0, $(V_{nn}+v_{ne})/t = -2$. Because $E_{total\ electr,0} = T+V_{ne}+V_{ee}+V_{nn}$ and $e_{total\ lectr,0} = t+v_{ne}+V_{nn}$, the virial theorem provides for atoms and stationary molecules that $E_{total\ electr,0} = -T$ and $e_{total\ electr,0} = -t$. While Eq.4 holds anywhere on the PES; the simpler form of virial theorem in Eq.22 is restricted to atoms and stationary points. As a consequence, Eq.4 can be expanded with the virial theorem with the restriction of Eq.22 for difference t-T as

$$t-T = E_{total\ electr,0} - e_{total\ electr,0} = E_{electr,0} - e_{electr,0} = \langle\Psi_0|H_{ee}|Y_0\rangle/\langle\Psi_0|Y_0\rangle = (N(N-1)/2)\xi \,. \qquad (23)$$

$E_{electr,0} - e_{electr,0} \gg 0$ (Eq.10) with Eq.23 yields $t \gg T$.

Important "back restrictions" of Eq.23 is that only atoms strictly obey Eq.23, the reason: If parameter "a" alters, the stationary geometries for t, T, or equivalently for $e_{electr,0}$ and $E_{electr,0}$, are not the same; but still the $t > T$ is true for any geometry. This argument via Eq.23 is valid not only for [0,1] but for the end points of any $[a_1,a_2]$.

**Generalization of 1st Hohenberg-Kohn theorem (1964) from a=1 to a=0 (TNRS) and general "a":**
Beside $\Psi_0 \Rightarrow \{\rho_0, E_{electr,0}, \text{all properties}\}$, the 1st HK theorem [3, 5], $\rho_0 \Rightarrow \{N, Z_A, R_A\} \Rightarrow H \Rightarrow \Psi_0 \Rightarrow \{E_{electr,0}, \text{all properties}\}$, provides that $\Psi_0 \Leftrightarrow H \Leftrightarrow H_\nabla + H_{ne} \Leftrightarrow Y_0$, finally, $Y_0(a=0) \Leftrightarrow H_{ne} \Leftrightarrow \Psi_0(a=1)$ or

$$\rho_0(\mathbf{r}_1, a_1) \quad \text{or} \quad y_0(a_1) \Leftrightarrow \rho_0(\mathbf{r}_1, a_2) \quad \text{or} \quad y_0(a_2) , \quad (24)$$

$$\rho_0(\mathbf{r}_1, a=0) \text{ from } H_\nabla + H_{ne} \Leftrightarrow E_{electr,0} \text{ from } H_\nabla + H_{ne} + H_{ee} . \quad (25)$$

The latter is very important in DFT practice (see Eqs.4, 5, 20). The generalization of a 2nd HK theorem is in fact trivial, since $H_\nabla + H_{ne} + aH_{ee}$ is linear not only for the a=1 (source of 2nd HK) but also for a≠1. The HK theorems for excited states are still problematic for any "a" not only for a=1.

**Particular functional link between a=1 and a=0 (TNRS) in ground state (k=0):** The LCAO coefficients do not differ significantly (at least not in the vicinity of stationary points), so we can assume ∃ $w(\mathbf{r}_1, \mathbf{r}_2, ... \mathbf{r}_N)$ r-symmetric to improve $\Psi_0 \approx S_0$ by $\Psi_0 = w(\mathbf{r}_1, \mathbf{r}_2, ... \mathbf{r}_N)Y_0$. Analytical form of w is not simple and unknown. For example, using $w = \Pi_{i=1...N} p(\mathbf{r}_i)$ with a high enough quality LCAO for p, the $Y_0$ becomes energetically better, but remains a single Slater determinant belonging to a better basis set, and this $(\Pi_{i=1...N} p(\mathbf{r}_i))Y_0$ still cannot totally reach $\Psi_0$, because of its single determinant nature. If w is good enough, $wY_0$ may approach $\Psi_0$ more efficiently than $S_0$. With basis set limit and general "a", the link with r-symmetric w is

$$y_0(a) = w(\mathbf{r}_1, \mathbf{r}_2, ... \mathbf{r}_N, a)Y_0 , \quad (26)$$

between two $y_0(a \neq 0)$, particularly between $y_0(a=1) = \Psi_0$ and $y_0(a=0) = Y_0$. Spin-orbit coordinates ($\mathbf{x}_i$) is not necessary in w, since the pre-calculated $Y_0(a=0)$ already contains it. Furthermore,

$$\rho_0(\mathbf{r}_1, a=1) = N \int |\Psi_0|^2 := N \int w^2 |Y_0|^2 = N \int w^2 (\int |Y_0| \, ds_1...ds_N) d\mathbf{r}_2...d\mathbf{r}_N \text{ vs. } \rho_0(\mathbf{r}_1, a=0) = N \int |Y_0|^2 ds_1 d\mathbf{x}_2...d\mathbf{x}_N \quad (27)$$

with normalization $\langle y_0(a)|y_0(a)\rangle = \langle wY_0|wY_0\rangle = 1 = \langle Y_0|Y_0\rangle$, and changing from $Y_0$ to $wY_0$ may need renormalization. A DFT version with $w_{DFT}(\mathbf{r}_1)$ between the physical and TNRS one-electron densities is

$$\rho_0(\mathbf{r}_1, a=1) = [w_{DFT}(\mathbf{r}_1)]^2 \rho_0(\mathbf{r}_1, a=0), \quad (28)$$

the square ensures $\rho_0 \geq 0$, the normalization is $N = \int \rho_0(\mathbf{r}_1, a=1) d\mathbf{r}_1 = \int [w_{DFT}(\mathbf{r}_1)]^2 \rho_0(\mathbf{r}_1, a=0) d\mathbf{r}_1$, see more in Appendix 1 and 2. Aside from $\langle w|w\rangle = \infty$, $wY_0$ must be well behaved x-anti-symmetric function with normalization constraint $\langle wY_0|wY_0\rangle = 1$. As a counterpart of Eqs.19-20, with a pre-calculated $Y_0$ at a=0 the variation equation for "a" (multiplying Eq.35 of Appendix 1 with $(wY_0)^*$ from the left and integrating) is

$$\text{enrg}_{electr,0}(a) = e_{electr,0} - (N/2)\langle wY_0|Y_0\nabla_1^2 w\rangle - N\langle wY_0|\nabla_1 Y_0 \nabla_1 w\rangle + a\langle wY_0|H_{ee}|wY_0\rangle , \quad (29)$$

wherein the equality holds with the minimizing r-symmetric w via VP. If a=0 ⇒ w(a=0)=1 and $\text{enrg}_{electr,0}(a=0) = e_{electr,0}$, if a=1 ⇒ $\text{enrg}_{electr,0}(a=1) = E_{electr,0}$, as well as if $\Psi_0 \approx Y_0$ i.e., w(a=1)≈1 crude approximation is taken, then Eq.29 reduces to Eq.20 (simply because $\nabla_1 w = 0$). Comparing Eq.29 to Eq.20 with a=1, the $N\{(\langle wY_0|Y_0\nabla_1^2 w\rangle/2 + \langle wY_0|\nabla_1 Y_0 \nabla_1 w\rangle\}$ converts (corrects) the $t + v_{ne}$ to $T + V_{ne}$.

## COMPUTATION PROPERTIES OF TNRS (a=0)

**The quasi-constant property of ξ** in Eq.4 is illustrated with $E_{electr,0}(G3)$ [8-9] of 149 ground state equilibrium neutral ($\Sigma Z_A = N$) molecules (max($Z_A$)<11, N≤68, order # increases with N). Fundamental property is that the TNRS (a=0) has similar LCAO coefficients to that of HF-SCF/basis/a with any "a" for the same molecular system on the same basis level, most importantly at a=1. Generally, LCAO coefficients are quasi-independent on "a". The electron-electron repulsion energy ($E_{electr,0} - e_{electr,0}$ in Eq.4) is quasi-linear w/r to the number of electron pairs N(N-1)/2, and $\text{enrg}_{electr,0}(a)$ is quasi-linear w/r to "a" (Fig.1). Finally, ξ is roughly a quasi-constant irrespective of molecular frames ($H_{ne}$ i.e. $\{R_A, Z_A\}_{A=1}^M$) and the number of electrons (N) in atoms and in equilibrium molecular systems, recall e.g. the (exact) value 2 of virial theorem. This quasi-constant cannot provide enough accuracy in particular calculations, because its fluctuation (Fig.2) with $H_{ne}$ causes error larger than CA: If it was a rigorous constant, $E_{electr,0}$ (a=1) could be extrapolated simply and directly from $e_{electr,0}(a=0)$ by Eq.20. The two curves, $E_{electr,0}(G3)$ and $e_{electr,0}$ run together like the same fingerprints as a function of molecular frame supporting that case a=0 has rich pre-information for a=1.

Via Eq.4, for (these 149 neutral) equilibrium molecules form the ratio

$$(E_{electr,0} - e_{electr,0})/e_{electr,0} = (E_{electr,0}/e_{electr,0}) - 1 = \langle \Psi_0|H_{ee}|Y_0\rangle / [\langle \Psi_0|Y_0\rangle \langle Y_0|H_\nabla + H_{ne}|Y_0\rangle] . \quad (30)$$

Inspecting Fig.2, recall the robust change in energies $E_{electr,0}$ and $e_{electr,0}$ as a function of nuclear frame (not plotted) and the definitely non-robust change in the interval $-0.4 < (E_{electr,0} - e_{electr,0})/e_{electr,0} < -0.3$ (standard dev. ≈0.04 which decreases with increasing N) of the lower curve plotted, as well as the ≈0.9 value of the upper curve proves Eq.20. In Eq.30 not the exact functions, but the $E_{electr,0}(G3)$ and $e_{electr,0}$(HF-SCF/basis/a=0) values are used for Fig.2. By Eq.20 or $\lim_{a \to 0} \Psi_0 \to Y_0$, Eq.30 reduces to

$$(E_{electr,0}(TNRS) - e_{electr,0})/e_{electr,0} = \langle Y_0|H_{ee}|Y_0\rangle / \langle Y_0|H_\nabla + H_{ne}|Y_0\rangle , \quad (31)$$

as well as

$$(E_{electr,0} - e_{electr,0})/(E_{electr,0}(TNRS) - e_{electr,0}) = \langle\Psi_0|H_{ee}|Y_0\rangle/[\langle\Psi_0|Y_0\rangle\langle Y_0|H_{ee}|Y_0\rangle] \ . \quad (32)$$

We have demonstrated the behavior of $(N(N-1)/2)\langle\Psi_0|r_{12}^{-1}|Y_0\rangle/\langle\Psi_0|Y_0\rangle$ in Eqs.1 and 4 introduced via $\xi$ as the counterpart of the similar but not the same quantity $V_{ee} \equiv (N(N-1)/2)\langle\Psi_0|r_{12}^{-1}|\Psi_0\rangle$. The $\xi N(N-1)/2$ is the exact difference between the ground state electronic energy ($E_{electr,0}$) of the real energy operator or Hamiltonian, $H_\nabla + H_{ne} + H_{ee}$, with a ground state wave function $\Psi_0$ and the ground state energy ($e_{electr,0}$) of energy operator $H_\nabla + H_{ne}$ with ground state eigenfunction $Y_0$, while $V_{ee}$ is the energy part electron-electron repulsion in the $E_{electr,0}$ value coming from the operator part $H_{ee}$.

**Quick power series estimation for ground state (k=0) from a=0 (TNRS) to a=1** is possible for example, because Eq.20 recovers the large part of energy difference ($E_{electr,0} - e_{electr,0}$), although still far from CA. One can rewrite every density functional as a function of the moments of the electron density [7], $\rho_0^n$, making sure the moments are complete. It allows to replace the functional analysis in DFT with a simple multivariate calculus, which is a huge formal advantage. It assumes that quantities can be written as a linear function of the moments, e.g. the Thomas-Fermi ($T \approx c_F \int \rho_0^{5/3} d\mathbf{r}_1$) or Weizsacker approximation ($T \approx (1/8)\int|\nabla_1\rho_0|^2/\rho_0 d\mathbf{r}_1$), Dirac formula ($V_{ee} \approx B_D \int \rho_0^{4/3} d\mathbf{r}_1$), Parr terms $c_{AB}(\int \rho_0^A d\mathbf{r}_1)^B$ with wisely chosen A and B keeping it scaling correct up to infinity for both, T and $V_{ee}$, as well as Kristyan's approximation $E_{corr} \approx k_c \langle S_0|H_\nabla|S_0\rangle + k_{ee}\langle S_0|H_{ee}|S_0\rangle$, wherein $a = 1 + k_{ee}$ corrects the HF-SCF/basis/a=1 calculation for $E_{electr,0}$. In the latter, $0 < |k_c|, k_{ee} < 0.01$ is quasi-universal for stationer/equilibrium nuclear frames. (If all energy correction is only attributed to electron–electron interaction ($k_c := 0$), the HF-SCF/6-31G**/a=0.99353272 improves the average deviation of HF-SCF/6-31G**/a=1 from 0.7851 h to 0.1255 h on average in approximating $E_{electr,0}$.) Common in these formulas that, the first main terms come from plausible assumptions and derivations, but secondary and higher terms (with empirical parameters) definitely necessary for CA, (e.g. the Thomas-Fermi alone for T fails to describe chemical bonds, etc.). A drawback is that increasing the power decreases the range of molecular systems in terms of accuracy, so moment expansion has not come before DFT correlation calculations; not being as fortunate as in DFT formulas wherein the expressions are more compact (and not sums, see e.g. generalized gradient approximations (GGA)).

With the idea of TNRS, Eq.20 can be corrected using the pre-calculated $t \equiv \langle Y_0|H_\nabla|Y_0\rangle$, $v_{ne} \equiv \langle Y_0|H_{ne}|Y_0\rangle$ and $z \equiv a\langle Y_0|H_{ee}|Y_0\rangle$ in an $L^{th}$ order power series ("a" is not to be confused with coefficients $a_j$'s) as

$$E_{electr,0} \approx e_{electr,0} + \Sigma_{j=1}^L (a_j t^j + b_j v_{ne}^j + c_j z^j) = E_{electr,0}(TNRS) + (a_1 t + b_1 v_{ne} + (c_1 - 1)z) + \Sigma_{j=2}^L (a_j t^j + b_j v_{ne}^j + c_j z^j) \ . \quad (33)$$

Moller-Plesset perturbation uses the excited states ($Y_k$) in the expansion ("vertical" algebraic way), while Eq.33 uses only $Y_0$ ("horizontal" algebraic way). $2^{nd}$ order (L=2) coefficients in Eq.33 (by least square fitting to 149 ground state G3 molecular energies to minimize the average absolute deviation) are

```
a1= -0.761233,      b1= -0.448435,      c1=  0.430207
a2=  2.270220E-004, b2= -5.068453E-005, c2=  1.678742E-004
```

while the $3^{rd}$ order (L=3) coefficients are

```
a1= -0.853118,      b1= -0.519268,      c1=  0.289831
a2=  5.224182E-004, b2= -1.321651E-004, c2=  6.744563E-004
a3= -2.026111E-007, b3= -2.221198E-008, c3= -4.823247E-007.
```

The average absolute deviation in h and % and the maximum absolute deviation in h from G3 values are

```
L=2 in Eq.33     : 1.615905 h or 1.02 %,  7.015398 h
L=3 in Eq.33     : 1.563234 h or 1.06 %,  7.270620 h
HF-SCF/STO-3G/a=1: 3.497650 h or 1.88 %, 11.976560 h.
```

The correlation effect in HF-SCF/basis/a=1 is indeed corrected by Eq.33 (about $3.5 \rightarrow 1.6$ or $12 \rightarrow 7$), but the L=$3^{rd}$ order does not yield much better improvement over L=$2^{nd}$ order expansion for $E_{corr}$, a typical problem of this method. More, larger L yields not-realistic (very small) values (with alternating sign) for coefficients. Negative $a_1$ and $|a_1|<1$ is necessary to subtract a part of kinetic energy away (T is targeted) because $t >> T$ (Eq.23), also for $b_1$ to keep the virial theorem hold, as well as $0 < c_1 < 1$ should be to be plausible with Eq.11. Approximate value of ratio in Eq.32 is 0.8-0.9 (Fig.2) showing plausible correspondence with $c_1$. While HF-SCF/STO-3G/a=1 is variational ($E_{corr} < 0$), the Eq.33 is not (positive and negative deviation from $E_{electr,0}$), a known fact in DFT. (For magnitude, $E_{corr} \approx c(N-1)$ in HF-SCF/basis/a=1, i.e. mainly and quasi-linearly depends on N in molecular systems, but it depends on $H_{ne}$ also and $-0.045 < c[h] < -0.035$ [10] for basis= STO-3G or 6-31G*; for large molecules, macroscopic media, crystals and metals it reads as $(\partial E_{corr}/\partial N)_{H_{ne}} \approx c$, but this crude estimation is below CA, however, weighting with partial charges, CA can be reached.)

The general linear moment expansion, e.g. for $V_{ee}(a=1)$ is $V_{ee} \approx C\int\rho_0^j d\mathbf{r}_1 + \Sigma_{j=1}^n c_j \int \rho_0^j d\mathbf{r}_1$, where term with C comes from plausible derivation, the others with $c_j$ are empirical from a fit. For example, Parr scaling correct series are $V_{ee} \approx C_J[\int \rho_0^{6/5} d\mathbf{r}_1]^{5/3} + \Sigma_{j=1}^n C_{xj}[\int \rho_0^{[1+1/(3j)]} d\mathbf{r}_1]^j$ and an algebraically similar for T, but these forms have problem described above. More powerful forms are based on Padé approximation, e.g. the two

$$V_{ee} \approx V_{ee}^{Padé-1} \equiv \int [(\Sigma_i a_i \rho_0^i)/(1+\Sigma_j b_j \rho_0^j)] d\mathbf{r}_1 \text{ or } V_{ee} \approx V_{ee}^{Padé-2} \equiv (\Sigma_i a_i \int \rho_0^i d\mathbf{r}_1)/(1+\Sigma_j b_j \int \rho_0^j d\mathbf{r}_1). \quad (34)$$

In the right in Eq.34, $\int \rho_0 d\mathbf{r}_1 = N$ is consistent with the normalization and analytical evaluation is possible with GTO, but numerical integration is necessary for STO, while the left one needs numerical integration in both basis sets, as well as the right equation can be linearized for parameter fit. In these formulas the $\rho_0 \approx \rho_0(\mathbf{r}_1,\text{HF-SCF/basis/a})$, Parr worked directly with a=1, but a=0 (TNRS) is also a choice to estimate for a=1.

**Speed of convergence in HF-SCF/basis/a** can be represented e.g. with hydrogen-fluorid: 1 step (a=0, TNRS) vs. 5 steps (a=1). HF-SCF/basis/a=0 needs only two steps, more exactly one, after setting up an initial guess for LCAO parameters, the eigensolver yields the $Y_0$ in the next step, irrespectively of molecular size, (in fact HF-SCF convergence is not needed at all, only one step engensolving). Starting with the commonly used Harris approximation for initial LCAO parameters for performing HF-SCF/basis/a=1 and finishing the convergence, or starting with LCAO from a converged one step HF-SCF/basis/a=0 and finishing the convergence for a=1, the final $E_{electr,0}$(HF-SCF/basis/a=1) and LCAO parameters will be strictly the same via the VP (Eq.21) kept by the subroutine (e.g. Gaussian 09). Generally, the HF-SCF/basis/a with a=0 can be achieved in basically one step for molecules of any size via a HF-SCF algorithm, while for a≠0, the number of convergence is always much more than one and increases with molecular size, as well as larger molecular systems may have problems such as break down in convergence in later steps, experienced since long in practice wherein a=1. This only one step (a=0) is a benefit if a good Eq.33 or 34 (both based on Eq.20) follows to finish the calculation for a=1. The commonly used Harris approximation makes a crude initial guess for one-electron density, $\rho_0(\mathbf{r}_1,a=1)$, using spherical atoms in a molecular frame, while TNRS yields the $\rho_0(\mathbf{r}_1,\text{HF-SCF/basis/a=0})$ which includes a crude estimation for chemical bond and density around the atoms in the molecule deformed from the atomic spherical shape in the particular molecular environment.

**Generalization of Koopmans' theorem (1934)** from HF-SCF/basis/a=1 can be done for a general "a" as follows. For a=1 it states [4] that according to the closed-shell HF-SCF theory, the first ionization energy of a molecular system is equal to the negative of the orbital energy of the highest occupied molecular orbital (HOMO). Seemingly it is trivial, but in practice, if a system is given by $H_{ne}$ and N, one does not have to make two HF-SCF/basis/a=1 calculations for an $(S_0, E_{electr,0}(\text{HF-SCF}))$ pair with N and N-1, and taking the energy difference for estimating ionization energy, but one calculation for N is enough, because one of its intrinsic energy values, of the HOMO, is about the same (but not exactly the same) as the difference in $E_{total\ electr,0}$(HF-SCF) for N and N-1. The equilibrium geometry, encapsulated in $H_{ne}$ and $H_{nn}$, differs slightly between N and N-1, because there is a shrink in the lowest lying doubly occupied MOs in $S_0$ if N is reduced to N-1 by the stronger effect of the nuclear frame (which slightly expands) if the number of electrons decreases, - Koopmans' theorem comes from a purely mathematical derivation in HF-SCF formalism. Here we introduce a similar mathematical situation: If a system is given by operator $H_{nn}$ or $H_{ne}$ and N, then a=0 determines a=1 "somehow" via the coupling strength parameter. After finishing algorithm HF-SCF/basis/a=0, there exists an algorithm transferring the energy to a=1, e.g. the crude Eq.20, or the more accurate Eqs.33-34, etc..

The generalization of Koopmans' theorem: It holds for any coupling strength parameter "a", the proof [4] is exactly the same. Moreover, it is trivial for a=0, because in $Y_0$ the MOs from Eq.2 do not change if N decreases to N-1, which is not the case if a≠0, more, it holds for open-shell systems as well if a=0. As an example of Ne, the ionization energy is $-f_5(a=1) = 0.54305$ h from HF-SCF/STO-3G/a=1, the accurate CI calculation for a=1 and the experimental values are 0.7946 h and 0.79248 h, resp.. The error from STO-3G basis set is large (0.79248 -0.54305= 0.24943), because this basis set is designed for energy differences and not for absolute values. On the other hand, the large $-f_5(a=0) = 10.22405$ h from a HF-SCF/STO-3G/a=0 comes from not involving $V_{ee}$ in TNRS (a=0), but importantly, Eq.20 re-corrects the value to 0.54305 as the $-f_5(a=1)$, the accurate back transfer up to 5 digits is accidental and originates from the now fortunate, rigid STO-3G basis set.

**The Hund's rule (1927) in relation to a=0 (TNRS)** must be commented. Hund's rules refer to a set of three rules, which are used to determine the term symbol that corresponds to the ground state of a multi-electron atom. It was first empirically established and then later proven in HF-SCF theory [4], but generally it has not yet been proven for Eq.1 only for HF-SCF/basis/a=1. The first rule states for a given electron configuration, that the term with maximum multiplicity has the lowest energy. Therefore, the term with lowest energy is also the term with maximum $S=\Sigma s_i$. It tells us something about how the electronic structure builds up as N increases; however, contradictions may arise.

We analyze how Eq.20 obeys Hund's rule. In relation to the coupling strength parameter "a", the case of a=0 manifests for Hund's rule, since it annihilates in eigenvalues ($e_{electr,k}$,$Y_k$) for any k≥0. (It means that in degenerate states, e.g., atomic p orbitals, the high spin fill up is energetically the same as the lower spin fill up.) Table 1 lists some ground state cases, HF- SCF/6-31G**/a=1 in 4$^{th}$ column obeys Hund's rule, i.e., all high spins are more stable (square brackets), 5$^{th}$ column with a=0 (TNRS) does not obey Hund's rule, (zeros in square brackets), and 6$^{th}$ column with the promising Eq.20 obeys Hund's rule again and close in values to 4$^{th}$ column (a=1). In the square brackets the important energy differences are between the high and low spin states of the same atom (negative sign means more stable). The order of the values for the energy gap between high and low spin states is in agreement between columns 4$^{th}$ and 6$^{th}$. X= number of convergence steps in HF-SCF and Y= sources of error in the column. The high spin CI calculations have also been listed in h for C, N and O atoms under the multiplicity 2S+1 in the 3$^{rd}$ column to compare to columns 4$^{th}$, 5$^{th}$ and 6$^{th}$. (Using STO-3G minimal basis set, this energy gap shown in square brackets for C, N and O atoms is 0.10881, 0.16447, 0.14233 h, resp. for both 4$^{th}$ and 6$^{th}$ columns, i.e., there is no difference between the two columns in these values. The reason is that the STO-3G basis set contains one branch of Gaussians and is not flexible enough to change the LCAO parameters in this respect, yet yields reasonable values; an overlap like this is characteristic of TNRS via HF-SCF/STO-3G/a=1 vs. HF-SCF/STO-3G/a=0 with Eq.20, not an accidental coincidence.) From an analytical point of view, Hund's rule applies if a≠0, but not if a=0; the energy gap between high and low spin states goes to zero if a→0 and Hund's rule annihilates in this respect.

## APPENDICES

**Appendix 1.** With an input a≠0 of interest and pre-calculated $Y_0$(a=0) substitute Eq.26 into Eq.1 and use $\nabla_1^2(wY_0)= w(\nabla_1^2 Y_0)+ 2\nabla_1 Y_0 \nabla_1 w+ Y_0 \nabla_1^2 w$. $H_{ne}$ formally disappears, but it is inherently included in $Y_0$ as

$$-(1/2)Y_0 \Sigma_{i=1}^N \nabla_i^2 w -\Sigma_{i=1}^N \nabla_i Y_0 \nabla_i w + aH_{ee}wY_0 = (enrg_{electr,0}(a)- e_{electr,0})wY_0 . \quad (35)$$

It provides the variation equation Eq.29 for r-symmetric w, $Y_0$ separates it term by term. The $Y_0$ has an exact Slater determinant form, all N! terms are different but algebraically equivalent, so it is enough to consider the first, or the diagonal. For example, for N=2 the $Y_0= |\alpha_1 f_1, \beta_2 f_2> = \alpha_1\beta_2 f_1 f_2 - \alpha_2\beta_1 f_1 f_2$ splits to N!=2 parts, and equality holds for the terms with the same spin parts, now both terms yield the same: $(1/2)(\nabla_1^2 w+ \nabla_2^2 w) –(\nabla_1 \ln f_1 \nabla_1 w+ \nabla_2 \ln f_2 \nabla_2 w) + aw/r_{12}= (enrg_{electr,0}(a) - e_{electr,0})w$ after dividing by the spatial part $f_1 f_2$. As a particular example, consider (non-relativistic calculation for) an atom (1≤Z≤18, M=a=1) with N=2 electrons: $Y_0$(a=0) contains $\phi_i(1s)= f_i=2Z^{3/2}\exp(-Z|\mathbf{r}_i|)$ with $\varepsilon_i=-Z^2/2$ in a.u. for i=1,2 (no basis set error), yielding the exact

$$-(1/2)(\nabla_1^2+\nabla_2^2)w +Z(\nabla_1|\mathbf{r}_1|\nabla_1 w+\nabla_2|\mathbf{r}_2|\nabla_2 w) +w/r_{12}= (E_{electr,0}+Z^2)w. \quad (36)$$

**Appendix 2.** The simplest "eigenvalue diff. equation" in Eq.1 comes from M=0, N=2, a=1 as $[(-\nabla_1^2-\nabla_2^2)/2 + r_{12}^{-1}]z= \varepsilon z$ with analytical solution [11] for the lowest ε value as $z(\mathbf{r}_1,\mathbf{r}_2)= \exp(r_{12}/2)$ with ε=-0.25. (Z:=0 in Eq.36 reduces to this "diff. equation" also.) Its spatial part shows why simple Hartree product z= exp($r_{12}$/2)≈ p($\mathbf{r}_1$)p($\mathbf{r}_2$) cannot be accurate, moreover, Slater determinant cannot be an analytic solution. The necessity of a correlation calculation manifests in the anti-symmetrized form $\Psi_0\equiv (\alpha_1\beta_2-\alpha_2\beta_1)\exp(r_{12}/2) \approx S_0 \equiv (\alpha_1\beta_2-\alpha_2\beta_1)$p($\mathbf{r}_1$)p($\mathbf{r}_2$). Approximating the spatial part with one Gaussian (STO-1G basis set) yields exp($r_{12}$/2)≈ c.exp($r_{12}^2$/2)= c.exp($r_1^2$/2)exp($r_2^2$/2)exp(-($x_1 x_2+y_1 y_2+z_1 z_2$)), where p($\mathbf{r}_i$):= exp($r_i^2$/2) for i=1,2 (recall the 1s type AOs), and the role of r-symmetric w in Eq.26 can be recognized as w($\mathbf{r}_1,\mathbf{r}_2$,a=1,N=2,M=0):= exp(-($x_1 x_2+y_1 y_2+z_1 z_2$)) acting as a correlation function. Its energy equivalent is the correlation energy, $E_{corr}$, from $[(-\nabla_1^2-\nabla_2^2)/2 +r_{12}^{-1}]$(exp($r_{12}$/2)- p($\mathbf{r}_1$)p($\mathbf{r}_2$)). Approximation only, because Gaussian basis set was involved, as well as for $E_{corr}$ normalized <|> integral average must be calculated, since p($\mathbf{r}_1$)p($\mathbf{r}_2$) is not an eigenfunction. The w accounts for the Fermi and Coulomb hole, the STO-GTO type difference exp($r_{12}$/2)-c.exp($r_{12}^2$/2) is responsible for the basis set error, which goes to basis set limit if many GTO are used (here $\Psi_0$ is approximated, but in practice it is the $S_0$), and exp($r_{12}$/2)-p($\mathbf{r}_1$)p($\mathbf{r}_2$) is responsible for the basis set and correlation error. In this simple example a quasi-accurate $E_{corr}$ can be evaluated, because the accurate wave function is known, but for physically important cases (N>1, M>0) in Eq.1, the $\Psi_0$ (a=1) is unknown. (Notice that exp($r_{12}$/2) is not well behaving since its integral over d$\mathbf{r}_1$d$\mathbf{r}_2$ is infinite.) More sophisticated model, called "uniform electron gas" (defined as a large N→∝ in a cube of volume V→∝, but finite ρ=N/V, throughout which there is a uniform spread of positive charge sufficient to make the system neutral) has led to serious correlation calculations. Furthermore, for real ground- and excited states as well as HF-SCF ground state one-electron density $\int \nabla_1^2 \rho(\mathbf{r}_1)d\mathbf{r}_1= 0$ holds, allowing tricky manipulations in correlation calculations [6].

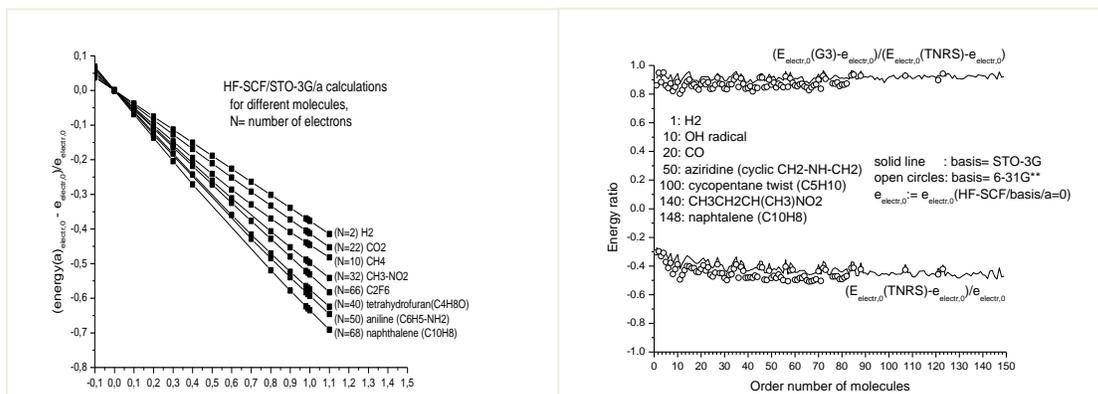

**FIGURE 1** (left): Plot of enrg$_{electr,0}$(a) in Eq.19 as quasi-linear function of coupling strength parameter "a". The slope at a=0 is $<Y_0|H_{ee}|Y_0>/<Y_0|H_\nabla+H_{ne}|Y_0>$ suffering from basis set error only, $y_0(a) \approx s_0(a)$ from HF-SCF/STO-3G/a, larger basis set yields slightly larger curvature (not shown).

**FIGURE 2** (right): $E_{electr,0}$(G3) includes $E_{corr}$ and correction for basis set error, but $e_{electr,0}$ from HF-SCF/basis/a=0 suffers from basis set error. The robust change in $E_{electr,0}$ or $e_{electr,0}$ as a function of nuclear frame (order #) has disappeared in this plot, and the quasi-constant character of ξ on nuclear frame follows as the small $H_2$ (N=2) and large naphthalene (N=68) have about the same energy ratios (using $\Psi_0$(a=1, $V_{ee}$ included) and TNRS $Y_0$(a=0, no $V_{ee}$) in Eqs.31-32). $V_{nn}$ is not included at all.

**TABLE 1**: Exhibiting how Hund's rule applies as coupling strength parameter "a" alters

| Atom | Configuration above [$1s^2 2s^2$] | 2S+1, $E_{electr,0}$(CI) | a=1, $E_{electr,0} \approx <S_0|H|S_0>$ | a=0 (TNRS) $e_{electr,0}$ | a=0, $E_{electr,0}$(TNRS) from Eq.20 |
|---|---|---|---|---|---|
| C | $2p_x 2p_y$ | 3, triplet, -37.8450 | -37.680860 [-0.09230] | -53.106285 [0] | -35.971284 [-0.14335] |
| C | $2p_x^2$ | 1, singlet | -37.588558 | -53.106285 | -35.827936 |
| N | $2p_x 2p_y 2p_z$ | 4, quadruplet, -54.5893 | -54.385442 [-0.13966] | -77.929276 [0] | -52.145336 [-0.22481] |
| N | $2p_x^2 2p_y$ | 2, dublet | -54.245778 | -77.929276 | -51.920527 |
| O | $2p_x^2 2p_y 2p_z$ | 3, triplet, -75.0674 | -74.783934 [-0.12733] | -109.338617 [0] | -71.698628 [-0.19431] |
| O | $2p_x^2 2p_y^2$ | 1, singlet | -74.656604 | -109.338617 | -71.504319 |
| X | - | - | 7-11 | 1 | 1 |
| Y | - | - | basis set, $E_{corr}$ | basis set | basis set, $E_{corr}$ |

## ACKNOWLEDGMENTS

Financial and emotional support for this research from OTKA-K 2015-115733 and 2016-119358 are kindly acknowledged. Special thanks to Hermin Szeger for her help in typing the manuscript. The subject has been presented in ICNAAM_2019, Greece, Rhodes and AIP 2020.